\documentclass[prb,twocolumn,showkeys,showpacs,amsmath,amsfonts,floatfix,superscriptaddress,nofootinbib]{revtex4-1}

\usepackage{amssymb,amsmath,amstext}                %%    American Physical Society math etc extensions
\usepackage{graphicx}                                               %%    Include figure files                                            %%   Manage links
\usepackage{epstopdf}                                               %%   help with eps -> pdf 
\usepackage{color}       
\usepackage{bm}% bold math                                              %%   change color
\usepackage{appendix}                                              %%   formating appendicies
\usepackage[latin2]{inputenc}
\usepackage{ulem}
\usepackage{latexsym}
\usepackage[colorlinks=true,citecolor=blue,linkcolor=magenta]{hyperref}
\def\be{\begin{equation}}
\def\ee{\end{equation}}
\def\bea{\begin{eqnarray}}
\def\eea{\end{eqnarray}}
\def\bi{\begin{itemize}}
\def\ei{\end{itemize}}

\begin{document}

\title{ Time Evolution of an Infinite Projected Entangled Pair State:\\
               an Algorithm from First Principles                          }

\author{ Piotr Czarnik }
\affiliation{Institute of Nuclear Physics, Polish Academy of Sciences, 
             Radzikowskiego 152, PL-31342 Krak\'ow, Poland}

\author{Jacek Dziarmaga}
\affiliation{Institute of Physics, Jagiellonian University, 
             {\L}ojasiewicza 11, PL-30348 Krak\'ow, Poland}

\date{\today}

\begin{abstract}
A typical quantum state obeying the area law for entanglement on an infinite 2D lattice can be represented by a tensor network ansatz -- known as an infinite projected entangled pair state (iPEPS) -- with a finite bond dimension $D$. Its real/imaginary time evolution can be split into small time steps. An application of a time step generates a new iPEPS with a bond dimension $k$ times the original one. The new iPEPS does not make optimal use of its enlarged bond dimension $kD$, hence in principle it can be represented accurately by a more compact ansatz, favourably with the original $D$. In this work we show how the more compact iPEPS can be optimized variationally to maximize its overlap with the new iPEPS. To compute the overlap we use the corner transfer matrix renormalization group (CTMRG). By simulating sudden quench of the transverse field in the 2D quantum Ising model with the proposed algorithm, we provide a proof of principle that real time evolution can be simulated with iPEPS. A similar proof is provided in the same model for imaginary time evolution of purification of its thermal states.
\end{abstract}

\maketitle

%%%%%%%%%%%%%%%%%%%%%%%%%%%%%%%%%%%%%%%%%%%%%%%%%%%%%%%%%%%%%%%%%%%%%%%%%%
\section{Introduction}
\label{sec:introduction}
%%%%%%%%%%%%%%%%%%%%%%%%%%%%%%%%%%%%%%%%%%%%%%%%%%%%%%%%%%%%%%%%%%%%%%%%%%

Tensor networks are a natural language to represent quantum states of strongly correlated systems\cite{Verstraete_review_08,Orus_review_14}. Among them the most widely used ansatze are a matrix product states (MPS) \cite{Fannes_MPS_92} and its 2D generalization: pair-entangled projected state (PEPS) \cite{Verstraete_PEPS_04} also known as a tensor product state. Both obey the area law for entanglement entropy.  
In 1D matrix product states are efficient parameterizations of ground states of gapped local Hamiltonians \cite{Verstraete_review_08,Hastings_GSarealaw_07,Schuch_MPSapprox_08} and purifications of thermal states of 1D local Hamiltonians \cite{Barthel_1DTMPSapprox_17}. MPS is the ansatz optimized by the density matrix renormalization group (DMRG) \cite{White_DMRG_92, White_DMRG_93} which is one of the most powerful methods to simulate not only ground states of 1D systems but also theirs exited states, thermal states or dynamic properties \cite{Schollwock_review_05,Schollwock_review_11}. 

PEPS  are expected to be an efficient parametrization of ground states of 2D gapped local Hamiltonians \cite{Verstraete_review_08,Orus_review_14} and were shown to be an efficient representation of thermal states of 2D local Hamiltonians \cite{Molnar_TPEPSapprox_15}, though in 2D there are limitations to the assumed representability of area-law states by tensor networks \cite{Eisert_TNapprox_16}. Furthermore tensor networks can be used to represent  efficiently systems with fermionic degrees of freedom \cite{Eisert_fMERA_09,Corboz_fMERA_09,Barthel_fTN_09,Gu_fTN_10} as was demonstrated for both finite \cite{Cirac_fPEPS_10}  and infinite PEPS \cite{Corboz_fiPEPS_10,Corboz_stripes_11}. 

PEPS was originally proposed as a varaitional ansatz for ground states of 2D finite systems \cite{Verstraete_PEPS_04,Murg_finitePEPS_07} generalizing earlier attempts  to construct trial wave-functions for specific 2D models using 2D tensor networks \cite{Nishino_2DvarTN_04}.  Efficient numerical methods enabling optimisation and controlled approximate contraction of infinite PEPS (iPEPS) \cite{Cirac_iPEPS_08,Xiang_SU_08,Gu_TERG_08,Orus_CTM_09} became basis for promising new methods for strongly correlated systems.   Among recent achievements of those methods are solution of a long standing magnetization plateaus problem in highly frustrated compound $\textrm{SrCu}_2(\textrm{BO}_3)_2$  \cite{Matsuda_SS_13,Corboz_SS_14} and obtaining coexistence of superconductivity and striped order in the underdoped regime of the  Hubbard  model -- a result  which is corroborated by other numerical methods (among them another tensor network approach - DMRG simulations of finite-width cylinders) -- apparently settling one of long standing controversies concerning that model \cite{Simons_Hubb_17}. Another example of a recent contribution of iPEPS-based methods to condensed matter physics is a problem of existence and nature of spin liquid phase in kagome Heisenberg antiferromagnet for which new evidence in support of gapless  spin liquid was obtained  \cite{Xinag_kagome_17}. This progress was accompanied and partly made possible by new developments in iPEPS optimization  \cite{Corboz_varopt_16,Vanderstraeten_varopt_16}, iPEPS contraction \cite{Fishman_FPCTM_17,Xie_PEPScontr_17,Czarnik_fVTNR_16},  energy extrapolations \cite{Corboz_Eextrap_16}, and  universality class estimation \cite{Corboz_FCLS_18,Rader_FCLS_18,Rams_xiD_18}. These achievements encourage attempts to use iPEPS to simulate  broad class of states obeying 2D area law like thermal states \cite{Czarnik_evproj_12,Czarnik_fevproj_14,Czarnik_SCevproj_15, Czarnik_compass_16, Czarnik_VTNR_15,Czarnik_fVTNR_16,Czarnik_eg_17,Dai_fidelity_17}, states of dissipative systems \cite{Kshetrimayum_diss_17} or exited states \cite{Vanderstraeten_tangentPEPS_15}.

Among alternative tensor network approaches to strongly correlated systems are methods of direct contraction and renormalization of a 3D tensor network representing a density operator of a 2D  thermal state \cite{Li_LTRG_11,Xie_HOSRG_12,Ran_ODTNS_12,Ran_NCD_13,Ran_THAFstar_18,Su_THAFoctakagome_17,Su_THAFkagome_17} and, technically challenging yet able to represent critical states with subleading logarithmic corrections to the area law, multi-scale entanglement renormalization ansatz (MERA) \cite{Vidal_MERA_07,Vidal_MERA_08}  and its generalization branching MERA \cite{Evenbly_branchMERA_14,Evenbly_branchMERAarea_14}.  Recent years brought also progress in using DMRG to simulate cylinders with finite width. Such simulations are routinely used alongside iPEPS to investigate 2D  systems ground states  (see e.g. Ref. \onlinecite{Simons_Hubb_17}) and were  applied recently also to  thermal states \cite{Stoudenmire_2DMETTS_17,Weichselbaum_Tdec_18}. 

In this work we test an algorithm to simulate either real or imaginary time evolution with iPEPS. The algorithm uses second order Suzuki-Trotter decomposition of the evolution operator into small time steps \cite{Trotter_59,Suzuki_66,Suzuki_76}. A straightforward application of a time step creates a new iPEPS with a bond dimension $k$ times the original bond dimension $D$. If not truncated, the evolution would result in an exponential growth of the bond dimension. Therefore, the new iPEPS is approximated variationally by an iPEPS with the original $D$. The algorithm is a straightforward construction directly from first principles with a minimal number of approximations controlled by the iPEPS bond dimension $D$ and the environmental bond dimension $\chi$ in CTMRG. It uses CTMRG \cite{Baxter_CTM_78,Nishino_CTMRG_96,Orus_CTM_09,Corboz_CTM_14} to compute fidelity between the new iPEPS and its variational approximation. The very calculation of fidelity between two close iPEPS was shown to be tractable only very recently \cite{Orus_GSfidelity_17}. In this work we go further and demonstrate that the fidelity can be optimized variationally effectively enough for  time evolution.
    
A challenging application of the method is real time evolution after a sudden quench. A sudden quench of a parameter in a Hamiltonian excites entangled pairs of quasiparticles with opposite quasimomenta that run away from each other crossing the boundary of the subsystem. Consequently, the number of pairs that are entangled across the boundary (proportional to the entanglement entropy) grows linearly with time requiring an exponential growth of the bond dimension. Therefore, a tensor network is doomed to fail after a finite evolution time. Nevertheless, matrix product states proved to be useful for simulating time evolution after sudden quenches in 1D \cite{Zaunerstauber_DPT_17}. As a proof of principle that the same can be attempted with iPEPS in 2D, in this work we simulate a sudden quench in the transverse field quantum Ising model. 

Moreover, there are other -- easier from the entanglement point of view -- potential applications of the real time variational evolution. For instance, a smooth ramp of a parameter in a Hamiltonian across a quantum critical point generates the entanglement entropy proportional to the area of the boundary times a logarithm of the Kibble-Zurek correlation length $\hat\xi$ that in turn is a power of the ramp time \cite{Cincio_KZ_07}. Thanks to this dynamical area law, the required $D$ instead of growing   exponentially with time saturates  becoming a power of the ramp time. Even stronger limitations may apply in many-body localization (MBL), where localized excitations are not able to spread the entanglement. Tensor networks have already been applied to  2D MBL phenomena \cite{Wahl_MBL_17}.  Finally, after vectorization of the density matrix, the unitary time evolution can be generalized to a Markovian master equation with a Lindblad superoperator, where local decoherence limits the entanglement making the time evolution with a tensor network feasible \cite{Montangero_master_16,Kshetrimayum_diss_17}.

Another promising application is imaginary time evolution generating thermal states of a quantum Hamiltonian. By definition, a thermal Gibbs state maximizes entropy for a given average energy. As this maximal entropy is the entropy of entanglement of the system with the rest of the universe, then -- by the monogamy of entanglement -- there is little entanglement left inside the system. In more quantitative terms, both thermal states of local Hamiltonians and iPEPS representations  of  density operators obey area law for mutual information making an iPEPS a good ansatz for thermal states \cite{Wolf_Tarealaw_08}. In this paper we evolve a purification of thermal states in the quantum Ising model obtaining results convergent to the variational tensor network renormalization (VTNR) introduced and applied to a number of models in \cite{Czarnik_VTNR_15,Czarnik_compass_16,Czarnik_fVTNR_16,Czarnik_eg_17}. This test is a proof of principle that thermal states can be obtained with the variational imaginary time evolution.

%%%%%%%%%%%%%%%%%%%%%%%%%%%%%%%%%%%%%%%%%%%%%%%%%%%%%%%%%%%%%%%%%%%%%%%%%%%%
\begin{figure}[h!]
\vspace{-0cm}
\includegraphics[width=0.9999\columnwidth,clip=true]{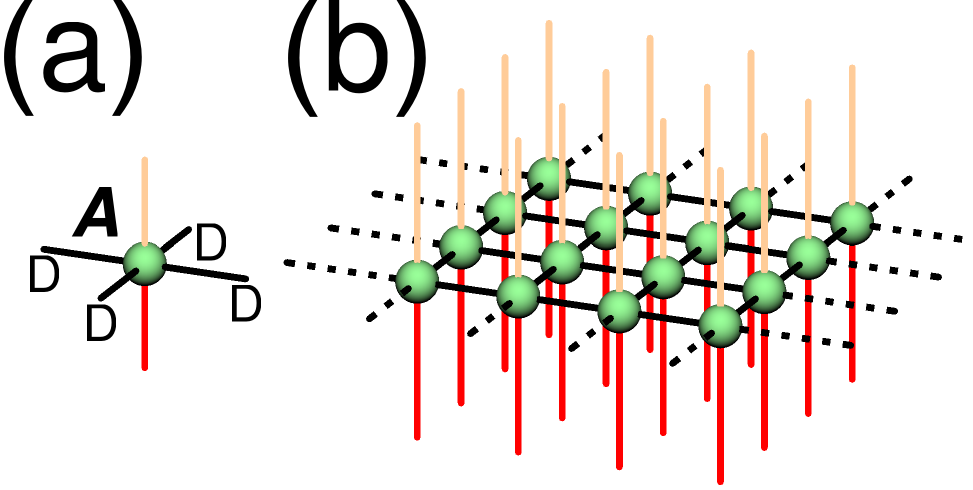}
\vspace{-0cm}
\caption{
In a,
an elementary rank-6 tensor $A$ of a purification. 
The top (orange) index numbers ancilla states $a=0,1$,
the bottom (red) index numbers spins states $s=0,1$,
the four (black) bond indices have a bond dimension $D$.
In b,
an iPEPS representation of the purification.
Here pairs of elementary tensors at NN lattice sites 
were contracted through their connecting bond indices.
The whole network is an amplitude for a joint spins' and ancillas' state 
labelled by the open spin and ancilla indices.
Reducing the dimension of ancilla indices to 1 (or simply ignoring the ancilla lines) 
we obtain a well known iPEPS representation of a pure state.   
}
\label{fig:A}
\end{figure}
%%%%%%%%%%%%%%%%%%%%%%%%%%%%%%%%%%%%%%%%%%%%%%%%%%%%%%%%%%%%%%%%%%%%%%%%%%%%

The paper is organized as follows.
In section \ref{sec:purification} we introduce purification of a thermal state to be evolved in imaginary time.
In section \ref{sec:algorithm} we introduce the algorithm in the more general case of imaginary time evolution of a thermal state purification. A modification to real time evolution of a pure state is straightforward.
In subsection \ref{sec:ST} we make Suzuki-Trotter decomposition of a small time step and represent it by a tensor network.
In subsection \ref{sec:step} we outline the algorithm
whose further details are refined in subsections \ref{sec:fom},\ref{sec:lu}, and appendix \ref{sec:2site}.
In section \ref{sec:im} the algorithm is applied to simulate imaginary time evolution generating thermal states. Its results are compared with VTNR.
In section \ref{sec:re} the real time version of the algorithm is tested in the challenging problem of time evolution after a sudden quench. 
Finally, we conclude in section \ref{sec:conclusion}.

%%%%%%%%%%%%%%%%%%%%%%%%%%%%%%%%%%%%%%%%%%%%%%%%%%%%%%%%%%%%%%%%%%%%%%%%%% 
\section{Purification of thermal states}
\label{sec:purification}
%%%%%%%%%%%%%%%%%%%%%%%%%%%%%%%%%%%%%%%%%%%%%%%%%%%%%%%%%%%%%%%%%%%%%%%%%% 

We will exemplify the general idea with the transverse field quantum Ising model on an infinite square lattice
\be 
H ~=~
- \sum_{\langle j,j'\rangle}Z_jZ_{j'}
- \sum_j \left( h_x X_j + h_z Z_j \right).
\label{calH}
\ee
Here $Z,X$ are Pauli matrices. At zero longitudinal bias, $h_z=0$, the model has a ferromagnetic phase with a non-zero spontaneous magnetization $\langle Z \rangle$ for sufficiently small transverse field $h_x$ and sufficiently large inverse temperature $\beta$. At $h_x=0$ the critical  $\beta$ is $\beta_0=-\ln(\sqrt{2}-1)/2\approx 0.441$ and at zero temperature the quantum critical point is $h_0=3.04438(2)$ \cite{Deng_QIshc_02}. 

In an enlarged Hilbert space, every spin with states $s=0,1$ is accompanied by an ancilla with states $a=0,1$. The space is spanned by states $\prod_j |s_j,a_j\rangle$, where $j$ is a lattice site. The Gibbs operator at an inverse temperature $\beta$ is obtained from its purification $|\psi(\beta)\rangle$ (defined in the enlarged space) by tracing out the ancillas,
\be
\rho(\beta) \propto
e^{-\beta H} =
{\rm Tr}_{\rm ancillas}|\psi(\beta)\rangle\langle\psi(\beta)|.
\label{rhobeta}
\ee
At $\beta=0$ we choose a product over lattice sites,
\be
|\psi(0)\rangle = \prod_j ~\sum_{s=0,1} |s_j,s_j\rangle,
\label{psi0}
\ee
to initialize the imaginary time evolution
\be
|\psi(\beta)\rangle~=~
e^{-\frac12\beta H}|\psi(0)\rangle~=~
U(-i\beta/2)|\psi(0)\rangle.
\label{psibeta}
\ee
The evolution operator $U(\tau)=e^{-i\tau H}$ acts in the Hilbert space of spins.
With the initial state (\ref{psi0}) Eq. (\ref{rhobeta}) becomes
\be
\rho(\beta) ~\propto~ U(-i\beta/2) U^\dag(-i\beta/2).
\label{UU}
\ee
Just like a pure state of spins,
the purification can be represented by a iPEPS, see Fig. \ref{fig:A}.

%%%%%%%%%%%%%%%%%%%%%%%%%%%%%%%%%%%%%%%%%%%%%%%%%%%%%%%%%%%%%%%%%%%%%%%%%%

%%%%%%%%%%%%%%%%%%%%%%%%%%%%%%%%%%%%%%%%%%%%%%%%%%%%%%%%%%%%%%%%%%%%%%%%%%%%
\begin{figure}[h!]
\vspace{-0cm}
\includegraphics[width=0.9999\columnwidth,clip=true]{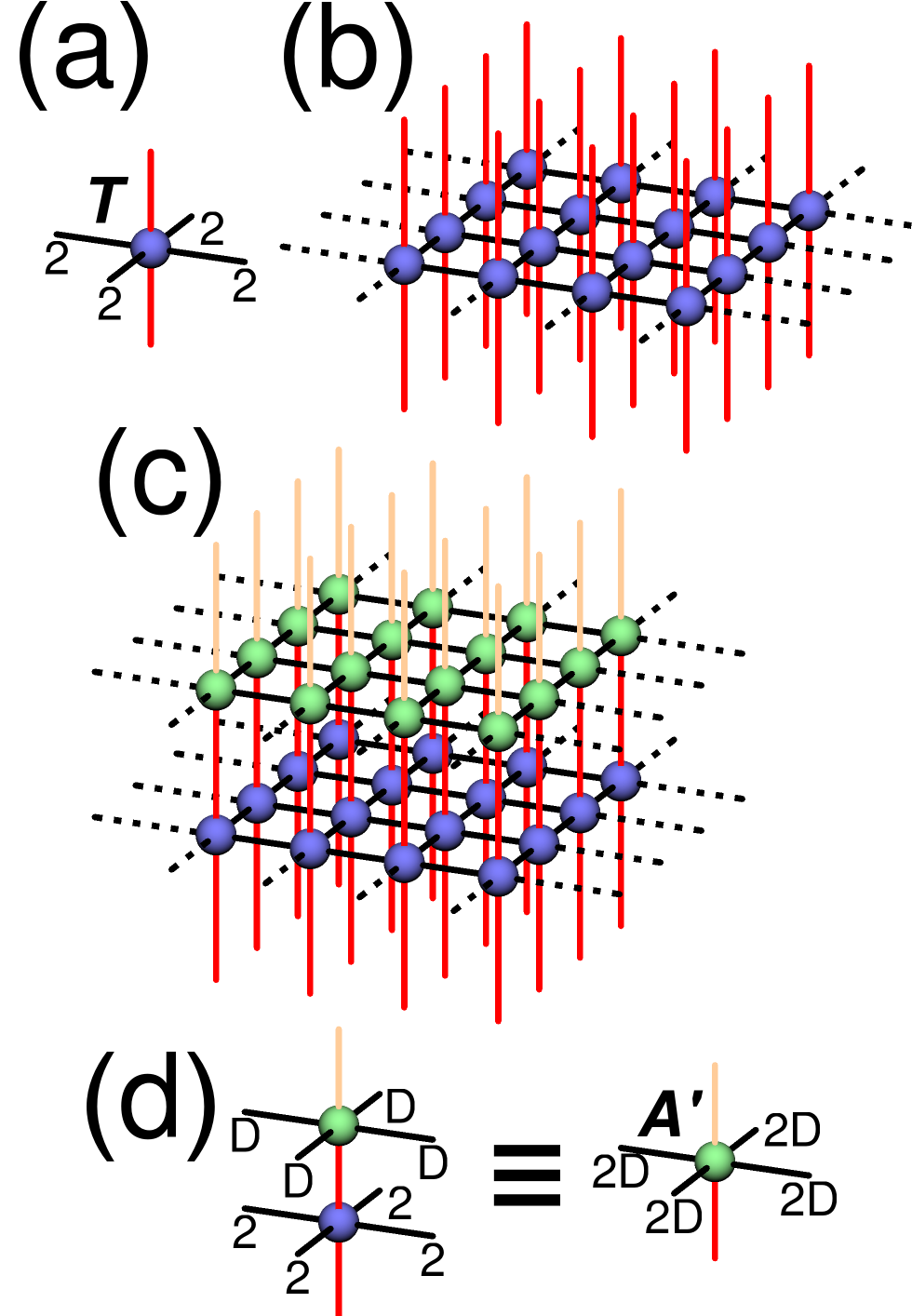}
\vspace{-0cm}
\caption{
In a,
an elementary rank-6 Trotter tensor $T$ with two (red) spin indices and four (black) bond indices, 
each of dimension 2.
In b,
a layer of Trotter tensors representing a small time step $U(d\tau)$.
In c,
the time step $U(d\tau)$ is applied to spin indices of the purification.
In d,
the tensors $T$ and $A$ can be contracted into a single new tensor $A'$.
A layer of $A'$ makes a new iPEPS that looks like the original one in Fig. \ref{fig:A}b but
has a doubled bond dimension $2D$.
}
\label{fig:AT}
\end{figure}
%%%%%%%%%%%%%%%%%%%%%%%%%%%%%%%%%%%%%%%%%%%%%%%%%%%%%%%%%%%%%%%%%%%%%%%%%%%%

%%%%%%%%%%%%%%%%%%%%%%%%%%%%%%%%%%%%%%%%%%%%%%%%%%%%%%%%%%%%%%%%%%%%%%%%%%
\section{The Method}
\label{sec:algorithm}
%%%%%%%%%%%%%%%%%%%%%%%%%%%%%%%%%%%%%%%%%%%%%%%%%%%%%%%%%%%%%%%%%%%%%%%%%%
We introduce the algorithm in the more general case of thermal states simulation by imaginary time evolution of their purification. To be more specific, we use the example of the quantum Ising model. Modification to real time evolution
amounts to ignoring any ancilla lines in the diagrams. For the sake of clarity, in the main text we fully employ 
the symmetry of the Ising model but we do our numerical simulations with a more efficient algorithm, described in 
Appendix \ref{sec:2site}, that breaks the symmetry by applying 2-site nearest-neighbor gates. That algorithm can be 
generalized to less symmetric models in a straightforward manner.

%%%%%%%%%%%%%%%%%%%%%%%%%%%%%%%%%%%%%%%%%%%%%%%%%%%%%%%%%%%%%%%%%%%%%%%%%%
\subsection{Suzuki-Trotter decomposition}
\label{sec:ST}
%%%%%%%%%%%%%%%%%%%%%%%%%%%%%%%%%%%%%%%%%%%%%%%%%%%%%%%%%%%%%%%%%%%%%%%%%%

In the second-order Suzuki-Trotter decomposition a small time step is
\bea
U(d\tau)
&=&
U_{h} (d\tau/2)
U_{ZZ}(d\tau  )
U_{h} (d\tau/2),
\label{Udbeta}
\eea
where 
\be
U_{ZZ}(d\tau) =
\prod_{\langle j,j'\rangle}e^{i d\tau Z_jZ_{j'}},~
U_{h}(d\tau)  =
\prod_j e^{i d\tau h_j}
\label{UZZ}
\ee
are elementary gates and $h_j = h_x X_j + h_z Z_j$.

In order to rearrange $U(d\tau)$ as a tensor network, 
we use singular value decomposition to
rewrite a 2-site term $e^{id\tau Z_jZ_{j'}}$ acting on a NN bond as a contraction of 2 smaller tensors
acting on single sites: 
\bea
e^{id\tau Z_jZ_{j'}} &=&
\sum_{\mu=0,1}
z_{j,\mu}
z_{j',\mu}.
\label{svdgate}
\eea
Here $\mu$ is a bond index and $z_{j,\mu}\equiv\sqrt{\Lambda_\mu}\,(Z_j)^\mu$ 
and  $\Lambda_0=\cos d\tau$ and $\Lambda_1=i\sin d\tau$.
Now we can write
\bea
U(d\tau)
&=&
\sum_{\{\mu\}}
\prod_j
\left[
e^{id\tau h_j/2}
\left(
\prod_{j'}
z_{j,\mu_{\langle j,j'\rangle}}
\right)
e^{id\tau h_j/2}
\right].
\label{Tx}
\eea
Here $\mu_{\langle j,j'\rangle}$ is a bond index on the NN bond $\langle j,j'\rangle$ 
and $\{\mu\}$ is a collection of all such bond indices. 
The square brackets enclose a Trotter tensor $T(d\tau)$ at site $j$, see Fig. \ref{fig:AT}a. 
It is a spin operator depending on the bond indices connecting its site with its four NNs.
A contraction of these Trotter tensors is the gate $U(d\tau)$ in
Fig. \ref{fig:AT}b. The evolution operator is a product of such time steps,
$U(Nd\tau)=\left[U(d\tau)\right]^N$.

%%%%%%%%%%%%%%%%%%%%%%%%%%%%%%%%%%%%%%%%%%%%%%%%%%%%%%%%%%%%%%%%%%%%%%%%%%%%
\begin{figure}[h!]
\vspace{-0cm}
\includegraphics[width=0.9999\columnwidth,clip=true]{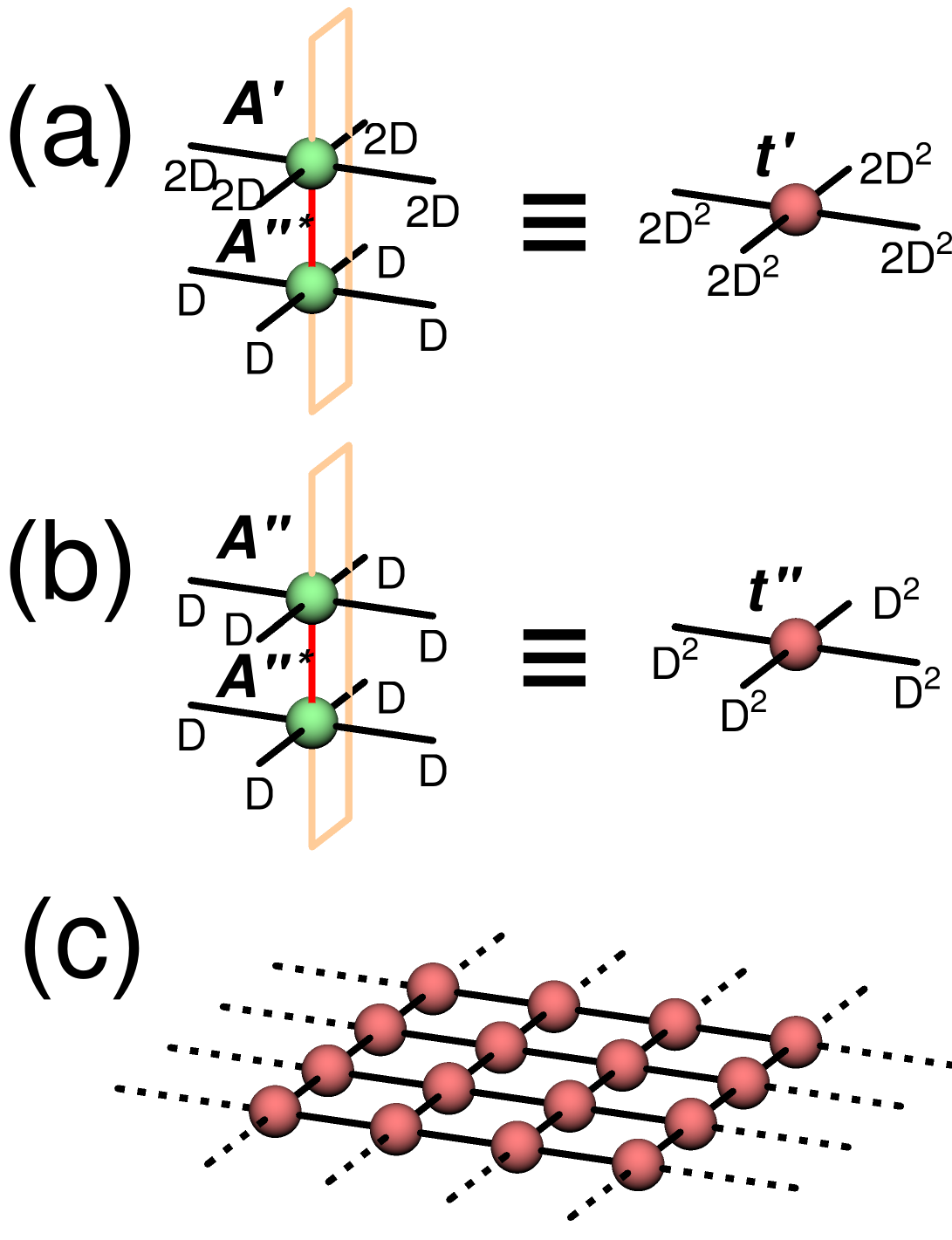}
\vspace{-0cm}
\caption{
In a,
tensor $A'$ is contracted with a complex conjugate of $A''$ 
into a transfer tensor $t'$ with a bond dimension $d=2D^2$.
In b,
tensor $A''$ is contracted with its own complex conjugate 
into a transfer tensor $t''$ with a bond dimension $d=D^2$.
In c,
an infinite layer of tensors $t'$ ($t''$)
represents the overlap $\langle\psi''|\psi'\rangle$ ($\langle\psi''|\psi''\rangle$).
}
\label{fig:t}
\end{figure}
%%%%%%%%%%%%%%%%%%%%%%%%%%%%%%%%%%%%%%%%%%%%%%%%%%%%%%%%%%%%%%%%%%%%%%%%%%%%

%%%%%%%%%%%%%%%%%%%%%%%%%%%%%%%%%%%%%%%%%%%%%%%%%%%%%%%%%%%%%%%%%%%%%%%%%%
\subsection{Variational truncation}
\label{sec:step}
%%%%%%%%%%%%%%%%%%%%%%%%%%%%%%%%%%%%%%%%%%%%%%%%%%%%%%%%%%%%%%%%%%%%%%%%%%

The time step $U(d\tau)$ applied to the state $|\psi\rangle$ yields a new state
\be 
|\psi'\rangle=U(d\tau)|\psi\rangle,
\ee
see Figs. \ref{fig:AT}c and d. 
If $|\psi\rangle$ has a bond dimension $D$,
then the new iPEPS has twice the original bond dimension $2D$.

In order to prevent exponential growth of the dimension in time, 
the new iPEPS has to be approximated by a more compact one, $|\psi''\rangle$, 
made of tensors $A''$ with the original bond dimension $D$.
The best $|\psi''\rangle$ minimizes the norm
\be 
\left| |\psi''\rangle - |\psi'\rangle \right|^2.
\label{norm}
\ee
Equivalently -- up to normalization of $|\psi''\rangle$ -- the quality of the approximation 
can be measured by a global fidelity
\be 
F=\frac{\langle\psi''|\psi'\rangle\langle\psi'|\psi''\rangle}{\langle\psi''|\psi''\rangle}.
\label{F}
\ee
After a rearrangement in section \ref{sec:fom} below, 
it becomes an efficient figure of merit.

The iPEPS tensor $A''$ -- the same at all sites -- has to be optimized globally.
However, the first step towards this global optimum is a local pre-update.
We choose a site $j$ and label the tensor at this site as $A''_j$.
This tensor is optimized while all other tensors are kept fixed as $A''$. 
With the last constraint the norm (\ref{norm}) becomes a quadratic form in $A''_j$.
The quadratic form is minimized with respect to $A''_j$ by $\tilde{A}$ that solves
the linear equation 
\be 
G\tilde A=V.
\label{GA=V} 
\ee
Here
\be
G=
\frac{\partial^2\langle\psi''|\psi''\rangle}{ \partial \left(A''_j\right)^* \partial \left(A''_j\right) },~~
V=
\frac{\partial \langle\psi''|\psi'\rangle}{\partial \left(A''_j\right)^* }
\label{GV}
\ee
are, respectively, a metric tensor and a gradient. 
Further details on the local pre-update can be found in section \ref{sec:lu} below.

The global fidelity (\ref{F}) is not warranted to increase when the local optimum $\tilde A$ is
substituted globally, i.e., in place of every $A''$ at every lattice site. However, $\tilde A$ 
can be used as an estimate of the most desired direction of the change of $A''$.
In this vein, we attempt an update
\be 
A''=A\cos\epsilon+\tilde A\sin\epsilon,
\label{Aepsilon}
\ee
with an adjustable parameter $\epsilon\in[-\pi/2,\pi/2]$ using an algorithm proposed in Ref. \onlinecite{Corboz_varopt_16} which simplified version was introduced in Refs. \onlinecite{Nishino_3Dstat_01,Gendiar_3Dstat_03}.
This update was successfully used in a similar variational problem of minimizing energy of an iPEPS as a function of $A$ \cite{Corboz_varopt_16}, where we refer for its detailed account. Here we just sketch the general idea.

To begin with, the global fidelity $F_0$ is calculated for the ``old'' tensor $A''=A$ with $\epsilon=0$.
For small $\epsilon$ the optimization is prone to get trapped in a local optimum. This is why large $\epsilon=\pi/2$ 
is tried first and if $F>F_0$ then $A''=\tilde A$ is accepted. Otherwise, $\epsilon$ is halved as many times as necessary 
for $F$ to increase above $F_0$ and then $A''=\tilde A$ is accepted. 
Negative $\epsilon$ are also considered in case the global $F$ does not increase for a positive $\epsilon$.

Once $A''$ in (\ref{Aepsilon}) is accepted, 
the whole procedure beginning with a solution of (\ref{GA=V}) is iterated until $F$ is converged.
The final converged $A''$ is accepted as a global optimum.

%%%%%%%%%%%%%%%%%%%%%%%%%%%%%%%%%%%%%%%%%%%%%%%%%%%%%%%%%%%%%%%%%%%%%%%%%%%%
\begin{figure}[h!]
\vspace{-0cm}
\includegraphics[width=0.9999\columnwidth,clip=true]{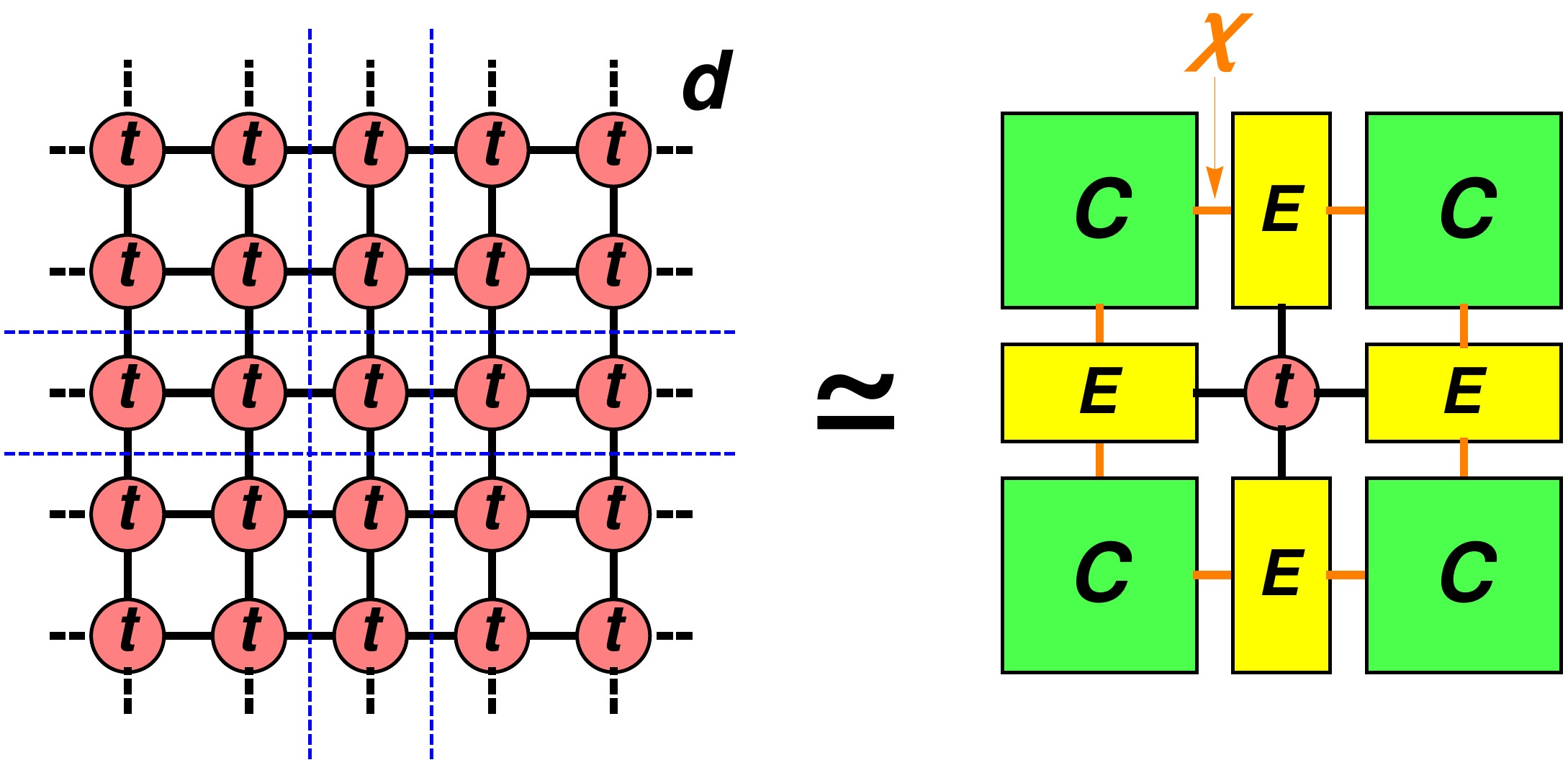}
\vspace{-0cm}
\caption{
Left, 
planar version of Fig. \ref{fig:t}c.
Right,
its approximate representation with corner tensors $C$ and edge tensors $E$.
Here $C$ effectively represents a corner of the infinite graph on the left
and $E$ its semi-infinite edge. Environmental bond dimension $\chi$ controls 
accuracy of the approximation. Tensors $C$ and $E$ are obtained with corner 
transfer matrix renormalization group \cite{Baxter_CTM_78,Nishino_CTMRG_96,Orus_CTM_09,Corboz_CTM_14}. 
}
\label{fig:CMR}
\end{figure}
%%%%%%%%%%%%%%%%%%%%%%%%%%%%%%%%%%%%%%%%%%%%%%%%%%%%%%%%%%%%%%%%%%%%%%%%%%%%

%%%%%%%%%%%%%%%%%%%%%%%%%%%%%%%%%%%%%%%%%%%%%%%%%%%%%%%%%%%%%%%%%%%%%%%%%%%%
\begin{figure}[h!]
\vspace{-0cm}
\includegraphics[width=0.8\columnwidth,clip=true]{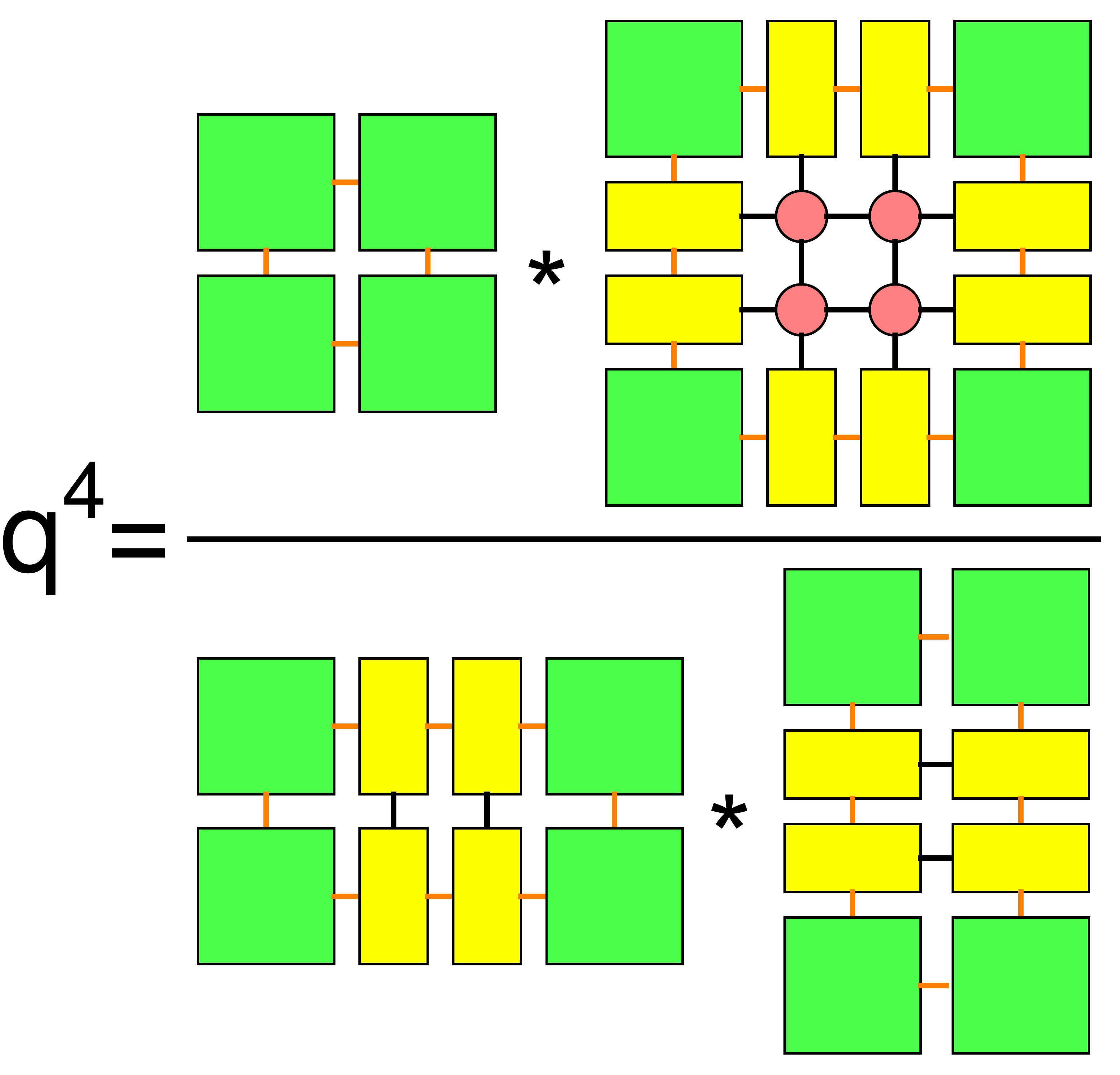}
\vspace{-0cm}
\caption{
The environmental tensors introduced in Fig. \ref{fig:CMR} can be used
to calculate the figure of merit (\ref{f}).  
This diagram shows a fourth power of a factor $q$ by which the diagram in Fig. \ref{fig:t}c
(or, equivalently, the left panel of Fig. \ref{fig:CMR})
is multiplied when $4$ sites are added to the network.
Depending on the overlap in question --
either $\langle\psi''|\psi'\rangle$ or $\langle\psi''|\psi''\rangle$,
see Fig. \ref{fig:t} -- 
the factor is either $n=q$ or $d=q$, respectively.
The diagram is equivalent to Fig.13.9 in R. J. Baxter's textbook
\cite{Baxter_Textbook_82}. 
}
\label{fig:Baxter}
\end{figure}
%%%%%%%%%%%%%%%%%%%%%%%%%%%%%%%%%%%%%%%%%%%%%%%%%%%%%%%%%%%%%%%%%%%%%%%%%%%%

%%%%%%%%%%%%%%%%%%%%%%%%%%%%%%%%%%%%%%%%%%%%%%%%%%%%%%%%%%%%%%%%%%%%%%%%%%
\subsection{Efficient fidelity computation}
\label{sec:fom}
%%%%%%%%%%%%%%%%%%%%%%%%%%%%%%%%%%%%%%%%%%%%%%%%%%%%%%%%%%%%%%%%%%%%%%%%%%

In the limit of infinite lattice, the overlaps in the fidelity (\ref{F}) become
\be
\langle\psi''|\psi'\rangle  = \lim_{N\to\infty} n^N,~~
\langle\psi''|\psi''\rangle = \lim_{N\to\infty} d^N,
\ee
where $N$ is the number of lattice sites. 
Consequently, the fidelity becomes
$
F = \lim_{N\to\infty} f^{N},
$ 
where 
\be 
f=\frac{n n^*}{d}
\label{f}
\ee
is a figure of merit per site. 

The factors $n$ and $d$ can be computed by CTMRG\cite{Orus_GSfidelity_17} generalizing the CTMRG approach to compute a partition function per site for 2D statistical models \cite{Baxter_CTM_78,Baxter_Textbook_82,Chan_fpsCTMRG_12,Chan_fpsCTMRG_13}. First of all, each overlap --
either $\langle\psi''|\psi'\rangle$ or $\langle\psi''|\psi''\rangle$ -- can be
represented by a planar network in Fig. \ref{fig:t}c. With the help of CTMRG \cite{Baxter_CTM_78,Nishino_CTMRG_96,Orus_CTM_09,Corboz_CTM_14}, 
this infinite network 
can be effectively replaced by a finite one, as shown in Fig. \ref{fig:CMR}. 
Figure \ref{fig:Baxter} shows how to obtain $n$ and $d$
with the effective environmental tensors introduced in Fig. \ref{fig:CMR}.

%%%%%%%%%%%%%%%%%%%%%%%%%%%%%%%%%%%%%%%%%%%%%%%%%%%%%%%%%%%%%%%%%%%%%%%%%%%%
\begin{figure}[h!]
\vspace{-0cm}
\includegraphics[width=0.9999\columnwidth,clip=true]{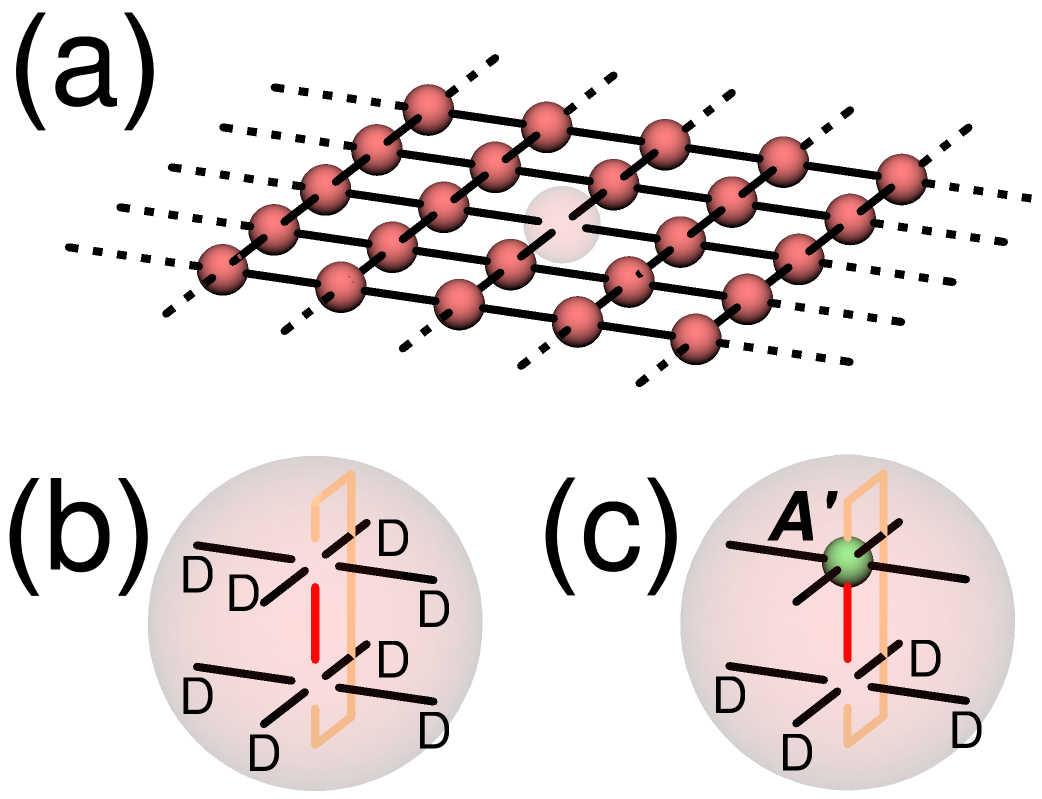}
\vspace{-0cm}
\caption{
In a,
tensor environment for $t'$ ($t''$). It is obtained by removing one tensor $t'$ ($t''$) from the overlap 
in Fig. \ref{fig:t}c or equivalently from the right diagram in Fig. \ref{fig:CMR}. 
The environment represents a derivative of the overlap $\langle\psi''|\psi'\rangle$ ($\langle\psi''|\psi''\rangle$)  with respect to $t'$ ($t''$)  (\ref{G2},\ref{V2}). This rank-4 tensor has 4 indices with dimension $D\times 2D$ ($D\times D$), 
respectively.
In b,
in case of $t''$ (\ref{G2}) each of the 4 indices in (a) is decomposed back into two indices, each of dimension $D$.
The diagram represents the metric tensor $G$.
The open (red) spin line is a Kronecker delta for spin states and the open (orange) ancilla line is a delta for ancillas.
Therefore, the metric can be decomposed as $G=g\otimes 1_s\otimes 1_a$, where $g$ is the tensor environment for $t$.
In c,
in case of $t'$ (\ref{V2}) each of the 4 indices in (a) is decomposed back into two indices of dimension $2D$ 
(upper) and $D$ (lower).
After contracting the upper indices with $A'$ the diagram becomes the gradient $V$.
}
\label{fig:GV}
\end{figure}
%%%%%%%%%%%%%%%%%%%%%%%%%%%%%%%%%%%%%%%%%%%%%%%%%%%%%%%%%%%%%%%%%%%%%%%%%%%%

%%%%%%%%%%%%%%%%%%%%%%%%%%%%%%%%%%%%%%%%%%%%%%%%%%%%%%%%%%%%%%%%%%%%%%%%%%
\subsection{Local pre-update}
\label{sec:lu}
%%%%%%%%%%%%%%%%%%%%%%%%%%%%%%%%%%%%%%%%%%%%%%%%%%%%%%%%%%%%%%%%%%%%%%%%%%
In order to construct $G$ and $V$ from the effective environmental tensors $C$ and $T$, 
it is useful to note first that a derivative of a contraction of two rank-$n$ tensors 
$f=\sum_{i_1,...,i_n} A_{i_1,...,i_n} B_{i_1,...,i_n}$ with respect to one of them gives 
the other one: $\partial f/\partial A_{i_1,...,i_n} = B_{i_1,...,i_n}$.
Futhermore,
we note that both the optimized tensor $A''_j$ and its conjugate $\left(A''_j\right)^*$ 
are located at the same site $j$ and they enter the overlap  
$\langle\psi''|\psi''\rangle$  ($\langle\psi''|\psi'\rangle$)  only through the 
tensor $t''$ ($t'$)defined in Fig. \ref{fig:t} (\ref{fig:t}a),
located at this site. We distinguish this tensor $t''$ ($t'$)  by an index $j$ and call it $t''_j$ ($t'_j$).
Therefore, the  derivatives in Eq. (\ref{GV}) decompose into a tensor contraction of
derivatives
\bea
G &=&
\frac{\partial\langle\psi''|\psi''\rangle}{\partial t''_j}
\frac{\partial^2 t''_j}{ \partial \left(A''_j\right)^* \partial \left(A''_j\right) }, 
\label{G2}\\
V &=&
\frac{\partial \langle\psi''|\psi'\rangle}{\partial t'_j}
\frac{\partial t'_j}{\partial \left(A''_j\right)^* }
\label{V2}
\eea
The derivatives of the overlaps with respect to $t'_j$ ($t''_j$) are represented by Fig. \ref{fig:GV}a, 
where one tensor $t'_j$ ($t''_j$) at site $j$ was removed from the overlap shown in Figs. \ref{fig:t}c, \ref{fig:CMR}. Indeed, a contraction of the missing $t'_j$ ($t''_j$) with its environment in Fig. \ref{fig:GV}a through corresponding
indices gives back the overlap. Diagramatically, this contraction amounts to filling the hole in Fig. \ref{fig:GV}a with the missing $t'_j$ ($t''_j$). In numerical calculations, the infinite diagram in Fig. \ref{fig:GV}a is approximated by a equivalent finite one in a similar way as in Fig. \ref{fig:CMR}.  

The rank-4 tensor in Fig. \ref{fig:GV}a is a tensor environment for $t'_j$ ($t''_j$).
Each of its 4 indices is a concatenation of two iPEPS bond indices, 
one from the ket and one from the bra iPEPS layer and 
has a dimension equal to ($2D\times D$) $D\times D$.
After splitting each index back into ket and bra indices, 
this environment can be used to calculate ($V$) $G$, 
as shown in Fig. \ref{fig:GV}c   (Fig. \ref{fig:GV}b). 
In Fig. \ref{fig:GV}b the hole in Fig. \ref{fig:GV}a (with split ket and bra indices) 
is filled with the second derivative of $t''_j$ with respect to $A''_j$ and $\left(A''_j\right)^*$. 
Similarly as the derivative of an overlap with respect to $t''_j$,
this derivative is obtained from the tensor $t''_j$ in Fig. \ref{fig:t}b by removing
both $A''_j$ and $\left(A''_j\right)^*$ from the diagram. 
In Fig. \ref{fig:GV}c the hole in Fig. \ref{fig:GV}a is filled by the derivative of $t'_j$ 
with respect to $\left(A''_j\right)^*$. 
This derivative is obtained from the tensor $t'_j$ in Fig. \ref{fig:t}a by removing
$\left(A''_j\right)^*$ from the diagram.

We have to keep in mind that the evironmental tensors are converged with limited precision 
that is usually set by demanding that local observables are converged with precision 
$\simeq 10^{-8}$. This precision limits the accuracy to which the matrix $G$ is Hermitean 
and positive definite. In order to avoid numerical instabilities this error has to filtered 
out by elliminating the anti-Hermitean part of $G$ and then truncating its eigenvalues that 
are less than a fraction of its maximal eigenvalue. The fraction is usually set at $10^{-8}$. 
To this end we solve the linear equation (\ref{GA=V}) using the Moore-Penrose pseudo-inverse 
\be 
\tilde{A}={\rm pinv}(G)V,
\label{tildeA}
\ee
where the truncation is implemented by setting an appropriate tolerance in the pseudo-inverse procedure.

Another advantage of the pseudo-inverse solution is that it does not contain any zero modes of $G$. 
By definition, these zero modes do not matter for the local optimization problem but 
they can make futile the attempt in (\ref{Aepsilon}) to use $\tilde{A}$ as a significant part
of the global solution. 

A possibility of further simplification occurs in Fig. \ref{fig:GV}b, where the open spin and ancilla lines represent two Kronecker symbols. The symbols are identities in the spin and ancilla subspace and, therefore, the metric $G$ has a convenient tensor-product structure $G=g\otimes 1_s\otimes 1_a$, where $g$ is a reshaped tensor environment for $t''_j$ and 
$1_s$ and $1_a$ are identities for spins and ancillas, respectively. 
Therefore -- after appropriate reshaping of tensors -- Eq. (\ref{tildeA}) can be reduced to
\be
\tilde A = {\rm pinv}(g) V,
\ee
where only the small tensor environment $g$ has to be pseudo-inverted.

%%%%%%%%%%%%%%%%%%%%%%%%%%%%%%%%%%%%%%%%%%%%%%%%%%%%%%%%%%%%%%%%%%%%%%%%%%%%
\begin{figure}[h!]
\vspace{-0cm}
\includegraphics[width=0.9999\columnwidth,clip=true]{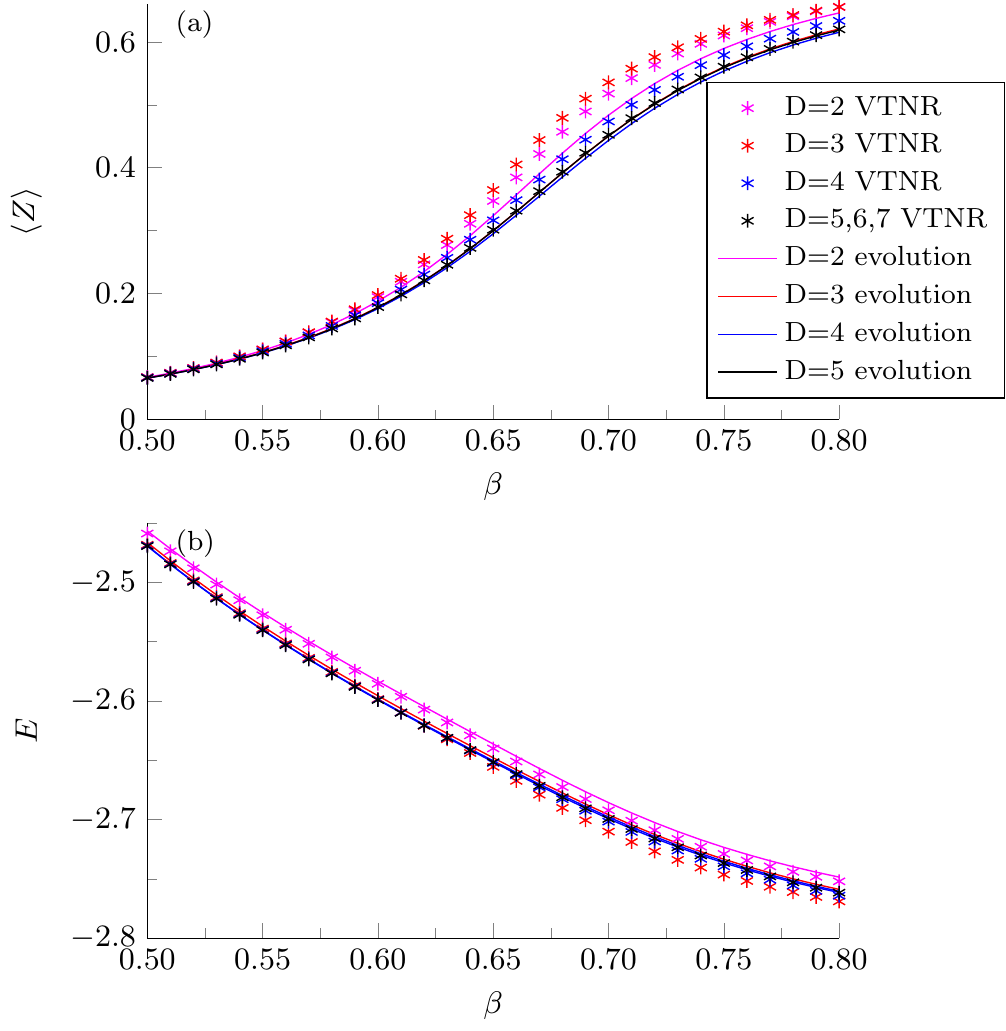}
\vspace{-0cm}
\caption{
Thermal states for a transverse field $h_x=2.5$ with a longitudinal bias $h_z=0.01$.
The stars are results from variational tensor network renormalization (VTNR)
and the solid lines from the imaginary time evolution.
With increasing bond dimension $D$ the two methods converge to each other.
In a, longitudinal magnetization  $\langle Z\rangle$ in function of inverse temperature.
In b, energy per site $E$ in function of inverse temperature.
}
\label{fig:imag25}
\end{figure}
%%%%%%%%%%%%%%%%%%%%%%%%%%%%%%%%%%%%%%%%%%%%%%%%%%%%%%%%%%%%%%%%%%%%%%%%%%%%
\begin{figure}[h!]
\vspace{-0cm}
\includegraphics[width=0.9999\columnwidth,clip=true]{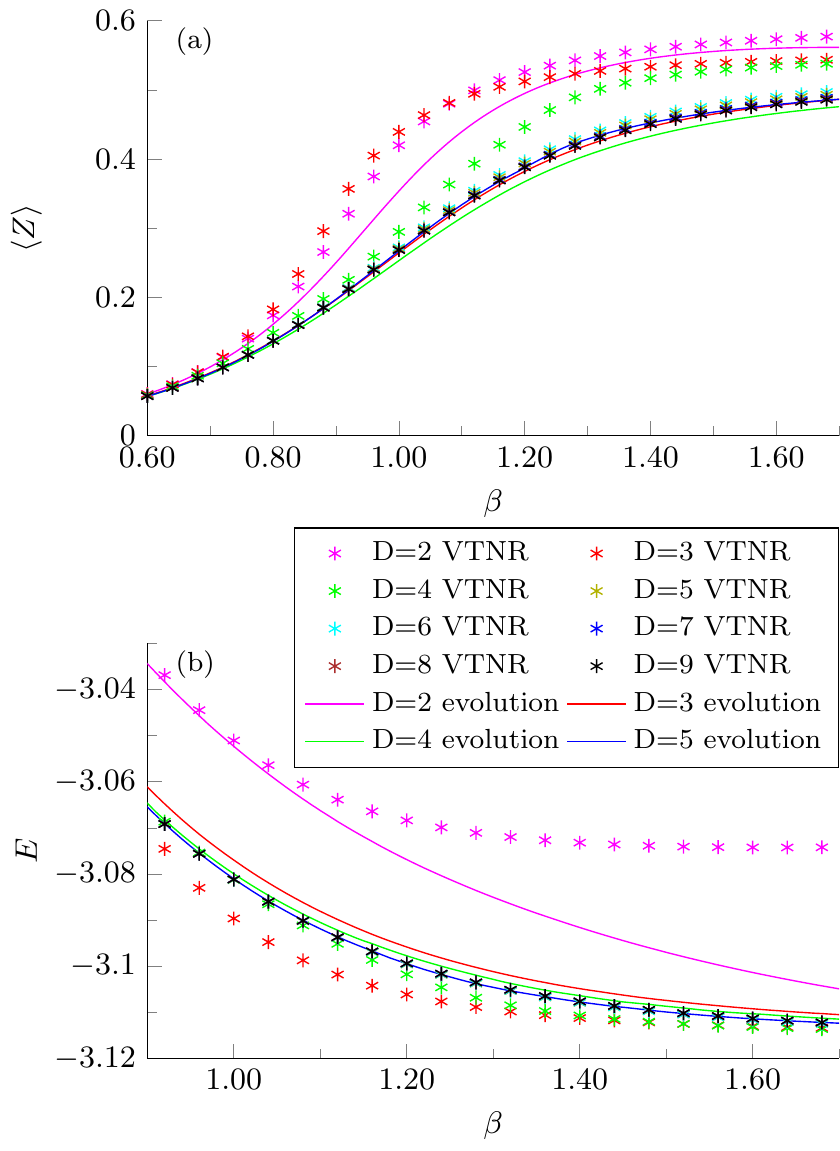}
\vspace{-0cm}
\caption{
Thermal states for a transverse field $h_x=2.9$ with a longitudinal bias $h_z=0.01$.
The stars are results from variational tensor network renormalization (VTNR)
and the solid lines from the imaginary time evolution.
With increasing bond dimension $D$ the two methods converge to each other.
In a, longitudinal magnetization $\langle Z\rangle$ in function of inverse temperature.
In b, energy per site $E$ in function of inverse temperature.
}
\label{fig:imag29}
\end{figure}
%%%%%%%%%%%%%%%%%%%%%%%%%%%%%%%%%%%%%%%%%%%%%%%%%%%%%%%%%%%%%%%%%%%%%%%%%%%%
%%%%%%%%%%%%%%%%%%%%%%%%%%%%%%%%%%%%%%%%%%%%%%%%%%%%%%%%%%%%%%%%%%%%%%%%%%
\section{Thermal states from imaginary time evolution}
\label{sec:im}
%%%%%%%%%%%%%%%%%%%%%%%%%%%%%%%%%%%%%%%%%%%%%%%%%%%%%%%%%%%%%%%%%%%%%%%%%%

In this section we present results obtained by imaginary time evolution for two values of 
the transverse field $h_x=2.5$ and $h_x=2.9$, see Figures \ref{fig:imag25} and \ref{fig:imag29}, 
corresponding to critical temperatures $\beta_c=0.7851(4)$ and $\beta_c=1.643(2)$, 
respectively \cite{Hesselmann_TIsingQMC_16}. 
We show data with $D=2,3,4,5$. 
The stronger field is closer to the quantum critical point at $h_0$, 
hence quantum fluctuations are stronger and a bigger bond dimension $D$ is required to converge. 
For the evolution to run smoothly across the critical point we added a small longitudinal bias $h_z=0.01$.

Figures \ref{fig:imag25}a and \ref{fig:imag25}b show the longitudinal magnetization 
$\langle Z\rangle$ and energy $E$ for the two transverse fields. 
The data from the evolution are compared to results obtained with the variational tensor network 
renormalization (VTNR) \cite{Czarnik_VTNR_15,Czarnik_compass_16,Czarnik_fVTNR_16,Czarnik_eg_17}. 
With increasing $D$ each of the two methods converges and they converge to each other.
This is a proof of principle that the variational time evolution can be applied to thermal states.   

The data at hand suggest that with increasing $D$ the evolution converges faster than VTNR.
However, at least for the Ising benchmark, 
numerical effort necessary to obtain results of similar accuracy is roughly the same.
In both methods the bottleneck is the corner transfer matrix renormalization procedure.
In the case of VTNR larger D is necessary  but in the case of the evolution the environmental tensors  
need to be computed more times.   

The advantage of VTNR is that it targets the desired temperature directly,
there is no need to evolve from $\beta=0$ and thus no evolution errors are accumulated.
In order to minimize the accumulation when evolving across the critical regime a small longitudinal bias
has to be applied. The critical singularity is recovered in the limit of small bias that
requires large $D$.
However, one big advantage of the variational evolution is that --
unlike VTNR targeting the accuracy of the partition function --
it aims directly at an accurate thermal state.
In some models this may prove to be a major advantage.

%%%%%%%%%%%%%%%%%%%%%%%%%%%%%%%%%%%%%%%%%%%%%%%%%%%%%%%%%%%%%%%%%%%%%%%%%%%%
\begin{figure}[h!]
\vspace{-0cm}
\includegraphics[width=0.9999\columnwidth,clip=true]{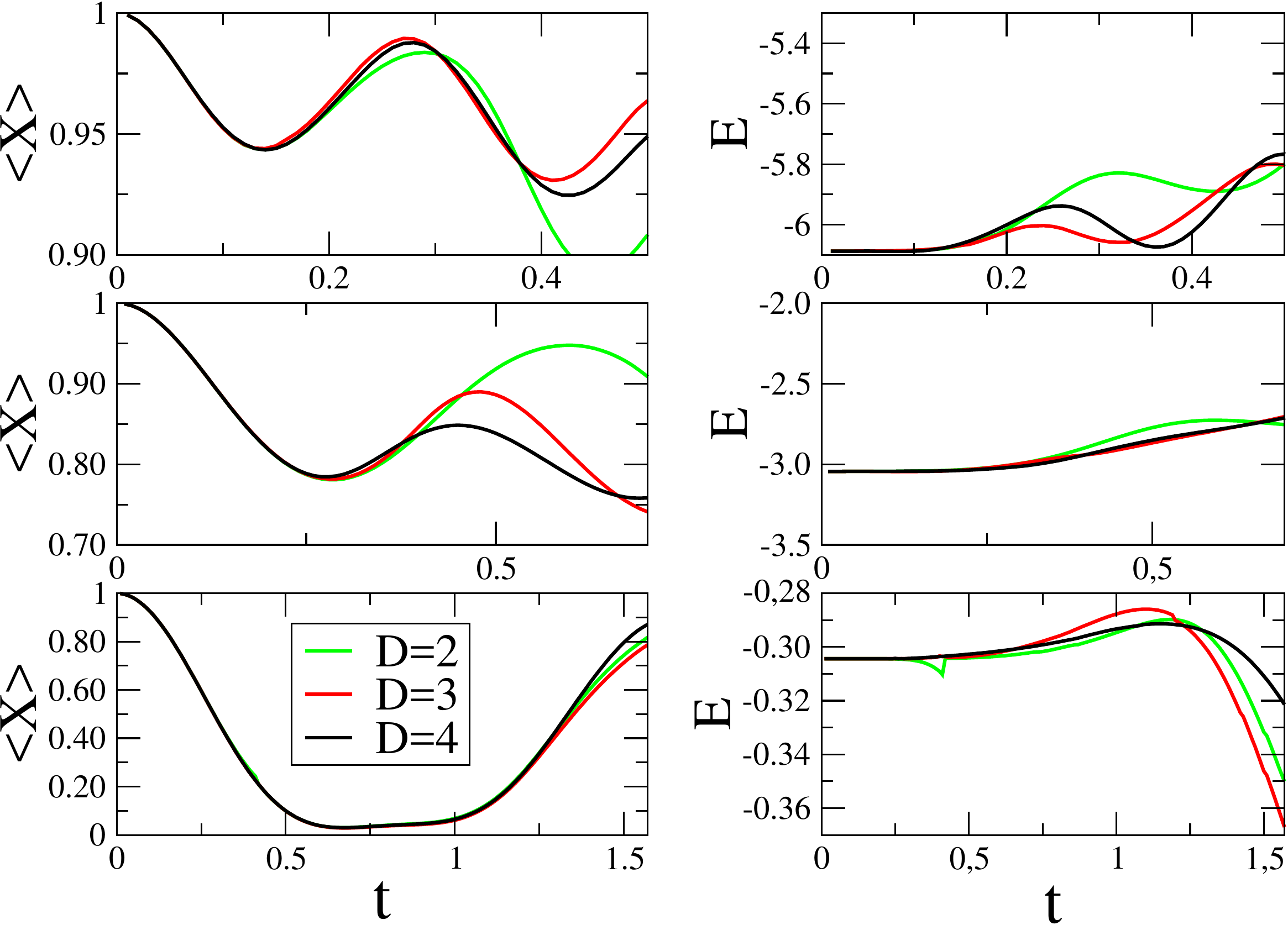}
\vspace{-0cm}
\caption{
Transverse magnetization $\langle X \rangle$ (left column) and 
energy per site (right column) after a sudden quench from a ground state
in a strong transverse field, $h_x\gg h_0$, with all spins pointing along $x$
down to a finite $h_x=2h_0$ (top row), $h_x=h_0$ (middle row), and $h_x=h_0/10$ 
(bottom row). The quench is, respectively, within the same phase, to the quantum 
critical point, and to a different phase.
Energy conservation shows systematic improvement with increasing
bond dimension $D=2,3,4$. We see that for sufficiently small times seemingly 
converged results for transverse magnetization can be obtained. While  
approaching the limit of small entanglement ($h_x=h_0/10$) we see that the 
"convergence time" is growing longer as expected.  
}
\label{fig:quench}
\end{figure}
%%%%%%%%%%%%%%%%%%%%%%%%%%%%%%%%%%%%%%%%%%%%%%%%%%%%%%%%%%%%%%%%%%%%%%%%%%%%
%%%%%%%%%%%%%%%%%%%%%%%%%%%%%%%%%%%%%%%%%%%%%%%%%%%%%%%%%%%%%%%%%%%%%%%%%%
\section{Time evolution after a sudden quench}
\label{sec:re}
%%%%%%%%%%%%%%%%%%%%%%%%%%%%%%%%%%%%%%%%%%%%%%%%%%%%%%%%%%%%%%%%%%%%%%%%%%

Next we move to simulation of a real time evolution after a quench in  an unbiased model (\ref{calH}) with $h_z=0$.  
The initial state is the ground state for $h_x\gg h_0$ with all spins 
pointing along $x$. At $t=0$ the Hamiltonian is suddenly quenched
down to a finite $h_x=2h_0,h_0,h_0/10$ that is, respectively, above, at, and below 
the quantum critical point $h_0$. 

Figure \ref{fig:quench} shows a time evolution of the magnetization 
$\langle X\rangle$ and energy per site $E$ after the sudden quench for bond 
dimensions $D=2,3,4$. With increasing $D$ the energy becomes conserved more
accurately for a longer time. This is an indication of the general convergence 
of the algorithm.

Not quite surprisingly, the results are most accurate for $h_x=h_0/10$. This weak transverse field is close to $h_x=0$ when the Hamiltonian is classical and the time evolution can be represented exactly with $D=2$. At $h_x=0$ quasiparticles have flat dispersion relation and do not propagate, hence -- even though they are excited as entangled pairs with opposite quasi-momenta -- they do not spread entanglement across the system. For any $h_x>0$, however, the entanglement grows with time and any bond dimension is bound to become insufficient after a finite evolution time. However, as discussed in Sec. \ref{sec:introduction}, there are potential applications where this effect is of limited importance.

%%%%%%%%%%%%%%%%%%%%%%%%%%%%%%%%%%%%%%%%%%%%%%%%%%%%%%%%%%%%%%%%%%%%%%%%%%
\section{Conclusion}
\label{sec:conclusion}
%%%%%%%%%%%%%%%%%%%%%%%%%%%%%%%%%%%%%%%%%%%%%%%%%%%%%%%%%%%%%%%%%%%%%%%%%%
We tested a straightforward algorithm to simulate real and imaginary time evolution with infinite iPEPS. The algorithm is based on variational maximization of a fidelity between a new iPEPS obtained after a direct application of a time step and its approximation by an iPEPS with the original bond dimension.

The main result is simulation of real time evolution after a sudden quench of a Hamiltonian. With increasing bond dimension the results converge over increasing evolution time. This is a proof of principle demonstration that simulation of a real time evolution with a 2D tensor network is feasible.

We also apply the same algorithm to evolve purification of thermal states. These results converge to the established VTNR method providing a proof of principle that the algorithm can be applied to 2D strongly correlated systems at finite temperature. 

%%%%%%%%%%%%%%%%%%%%%%%%%%%%%%%%%%%%%%%%%%%%%%%%%%%%%%%%%%%%%%%%%%%%%%%%%%%%%%
\acknowledgments
%%%%%%%%%%%%%%%%%%%%%%%%%%%%%%%%%%%%%%%%%%%%%%%%%%%%%%%%%%%%%%%%%%%%%%%%%%%%%%
P. C. acknowledges inspiring discussions with Philippe Corboz on application of CTMRG to calculation of partition function per site and simulations of thermal states.  We thank Stefan Wessel for numerical values of data publised in Ref. \onlinecite{Hesselmann_TIsingQMC_16}. Simulations were done with extensive use of ncon function \cite{ncon}. This research was funded by National Science Center, Poland under project 2016/23/B/ST3/00830 (PC) and QuantERA program 2017/25/Z/ST2/03028 (JD).

%%%%%%%%%%%%%%%%%%%%%%%%%%%%%%%%%%%%%%%%%%%%%%%%%%%%%%%%%%%%%%%%%%%%%%%%%%%%%%
\appendix
%%%%%%%%%%%%%%%%%%%%%%%%%%%%%%%%%%%%%%%%%%%%%%%%%%%%%%%%%%%%%%%%%%%%%%%%%%%%%%
\label{sec:appendix}
%%%%%%%%%%%%%%%%%%%%%%%%%%%%%%%%%%%%%%%%%%%%%%%%%%%%%%%%%%%%%%%%%%%%%%%%%%%%
\begin{figure}[h!]
\vspace{-0cm}
\includegraphics[width=0.9999\columnwidth,clip=true]{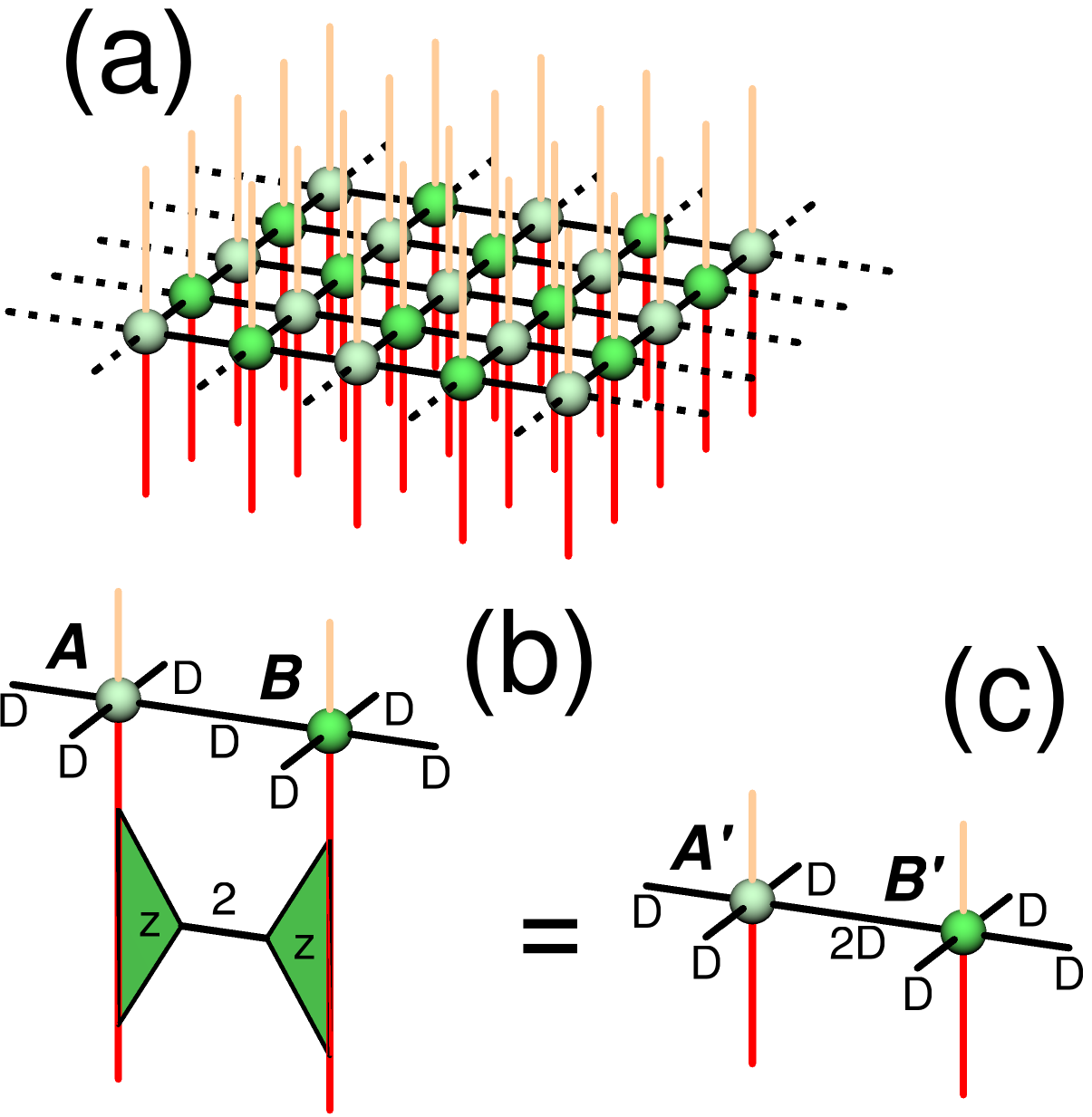}
\vspace{-0cm}
\caption{
In a,
the infinite square lattice is divided into two sublattices with tensors $A$ (lighter green) and $B$ (darker green). 
In b,
SVD decomposition of a NN gate is applied to every pair $A$ and $B$ of NN tensors.
In c,
when the tensors $A$ and $B$ are contracted with their respective $z$'s, 
then they become new tensors $A'$ and $B'$ with a doubled bond dimension $2D$ on their common NN bond. 
By variational optimization the iPEPS made of $A'$ and $B'$ is approximated
by a new iPEPS made of $A''$ and $B''$ with the original bond dimension $D$.  
}
\label{fig:2site}
\end{figure}
%%%%%%%%%%%%%%%%%%%%%%%%%%%%%%%%%%%%%%%%%%%%%%%%%%%%%%%%%%%%%%%%%%%%%%%%%%%%

%%%%%%%%%%%%%%%%%%%%%%%%%%%%%%%%%%%%%%%%%%%%%%%%%%%%%%%%%%%%%%%%%%%%%%%%%%
\section{2-site gates}
\label{sec:2site}
%%%%%%%%%%%%%%%%%%%%%%%%%%%%%%%%%%%%%%%%%%%%%%%%%%%%%%%%%%%%%%%%%%%%%%%%%%
For the sake of clarity, the main text presents a straightforward single-site version of the algorithm. 
In practice it is more efficient to implement the gate $U_{ZZ}(d\tau)$ as a product of two-site gates. 
To this end the infinite square lattice is divided into two sublattices $A$ and $B$, see Fig. \ref{fig:2site}a. 
On the checkerboard the gate becomes a product
\bea
&& 
U_{ZZ}(d\tau) = U^a_0(d\tau) U^a_1(d\tau) U^b_0(d\tau) U^b_1(d\tau).
\label{UZZ2s}
\eea
Here $a$ and $b$ are the Cartesian lattice directions spanned by $e_a$ and $e_b$,
\bea
U^a_s(d\tau)  &=&  \prod_{mn} e^{i d\tau Z_{2m+s-1,n}Z_{2m+s,n}},\label{Ua}\\
U^b_s(d\tau)  &=&  \prod_{mn} e^{i d\tau Z_{m,2n+s-1}Z_{m,2n+s}},\label{Ub}
\eea
and $Z_{m,n}$ is an operator at a site $me_a+ne_b$.

Every NN gate in (\ref{Ua},\ref{Ub}) is decomposed as in (\ref{svdgate}). 
Consequently, when a gate, say, $U^a_0(d\tau)$ is applied to the checkerboard $AB$-iPEPS in 
Fig. \ref{fig:2site}a, 
then every pair of tensors $A$ and $B$ at every pair of NN sites $(2m-1)e_a+ne_b$ and 
$2me_a+ne_b$ is applied with the NN-gate's decomposition as in Fig. \ref{fig:2site}b. 
When the tensors $A$ and $B$ are fused with their respective $z$'s, 
they become $A'$ and $B'$, respectively, that are connected by an index with a bond dimension $2D$, see Fig. \ref{fig:2site}c.
The action of the gate $U^a_0(d\tau)$ is completed when the $A'B'$-iPEPS is approximated by a (variationally optimized) new $A''B''$-iPEPS with the original bond dimension $D$ at every bond. 

Apart from the opportunity to use reduced tensors in the variational optimization, 
the main advantage of the 2-site gates is that the enlarged bond dimension $2D$ appears
only on a minority of bonds. This speeds up the CTMRG for the overlap $\langle\psi'|\psi''\rangle$
that is the most time-consuming part of the algorithm.
The decomposition into 2-site gates breaks the symmetry of the lattice. Therefore we use
the efficient non-symmetric version of CTMRG \cite{Corboz_CTM_14} for checkerboard lattice.

%%%%%%%%%%%%%%%%%%%%%%%%%%%%%%%%%%%%%%%%%%%%%%%%%%%%%%%%%%%%%%%%%%%%%%%%%%%%

%merlin.mbs apsrev4-1.bst 2010-07-25 4.21a (PWD, AO, DPC) hacked
%Control: key (0)
%Control: author (8) initials jnrlst
%Control: editor formatted (1) identically to author
%Control: production of article title (-1) disabled
%Control: page (0) single
%Control: year (1) truncated
%Control: production of eprint (0) enabled
%

\end{document}